\documentclass[conference]{IEEEtran}
\usepackage{cite}
\usepackage{amsmath,amssymb,amsfonts}
\usepackage{algorithmic}
\usepackage{graphicx}
\usepackage{textcomp}
\usepackage{xcolor}
\usepackage{hyperref}
\usepackage{subfig}
\usepackage{tabularx}
\usepackage{dblfloatfix}
\usepackage{xurl}

\begin{document}

\title{Fair Queuing Aware Congestion Control}

\author{\IEEEauthorblockN{Maximilian Bachl}
\IEEEauthorblockA{{\scriptsize \url{https://github.com/muxamilian/fair-queuing-aware-congestion-control}}}
}

\maketitle
\thispagestyle{plain}
\pagestyle{plain}

\begin{abstract}
Fair queuing is becoming increasingly prevalent in the internet and has been shown to improve performance in many circumstances. 
Performance could be improved even more if endpoints could detect the presence of fair queuing on a certain path and adjust their congestion control accordingly. 
If fair queuing is detected, the congestion control would not have to take cross traffic into account, which allows for more flexibility. 
In this paper, we develop the first algorithm that continuously checks if fair queuing is present on a path, with an accuracy of over 95\%. 
When fair queuing is detected, a different congestion control can be chosen, which can result in reduced latency. Also, each flow can then specify how much queuing delay it allows, meaning that it can choose its own tradeoff between throughput and latency. 
\end{abstract}

% \begin{IEEEkeywords}
% component, formatting, style, styling, insert
% \end{IEEEkeywords}

\begin{figure}[h]
\centering 
\subfloat[\footnotesize A host sending data to a receiver via two concurrent flows. The blue flow sends two times more data. There is a bottleneck link with a switch which applies \textbf{fair queuing}.]{\includegraphics[width=\columnwidth]{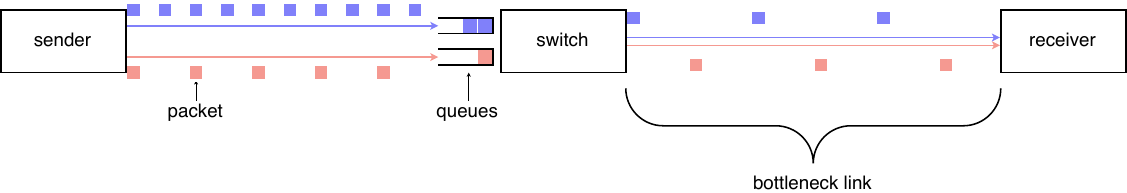}
\label{fig:fq_overview}}\\
\subfloat[\footnotesize Queuing delay in case of \textbf{fair queuing}.]{
\includegraphics[width=\columnwidth]{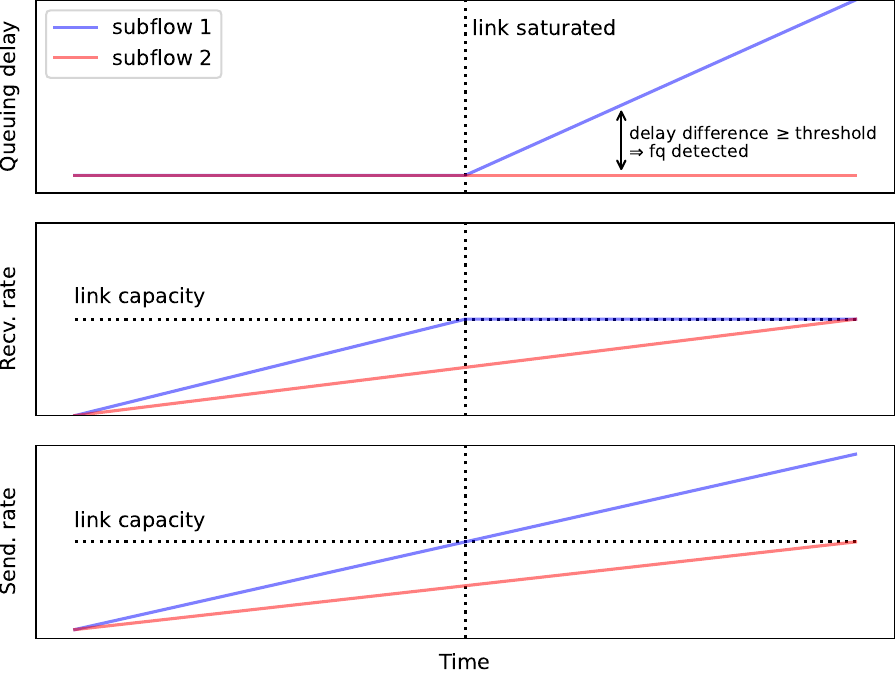}
\label{fig:fq_detail}}
\caption{In case there is \textbf{fair queuing}, when the bottleneck link is congested, subflow 1, which sends more, sees increased queuing delay. Subflow 2 doesn't see increased queuing delay as it sends less than its fair share and thus its packets don't have to wait at the bottleneck link.}
\label{fig:illustration_fq}
\end{figure}

\begin{figure}[h]
\centering
\subfloat[\footnotesize A host sending data to a receiver via two concurrent flows. The blue flow sends two times more data. There is a bottleneck link with a switch which applies \textbf{no fair queuing}.]{\includegraphics[width=\columnwidth]{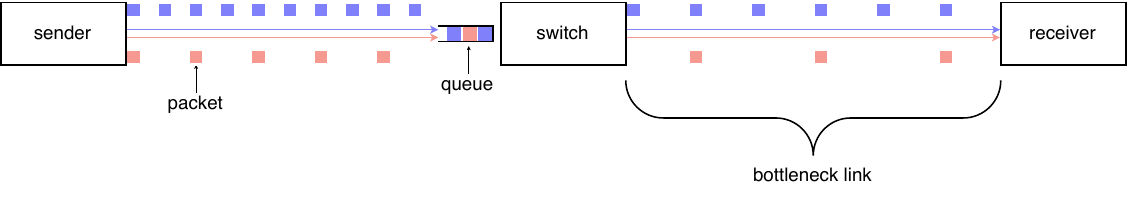}
\label{fig:no_fq_overview}}\\
\subfloat[\footnotesize Queuing delay in case there is \textbf{no fair queuing}.]{
\includegraphics[width=\columnwidth]{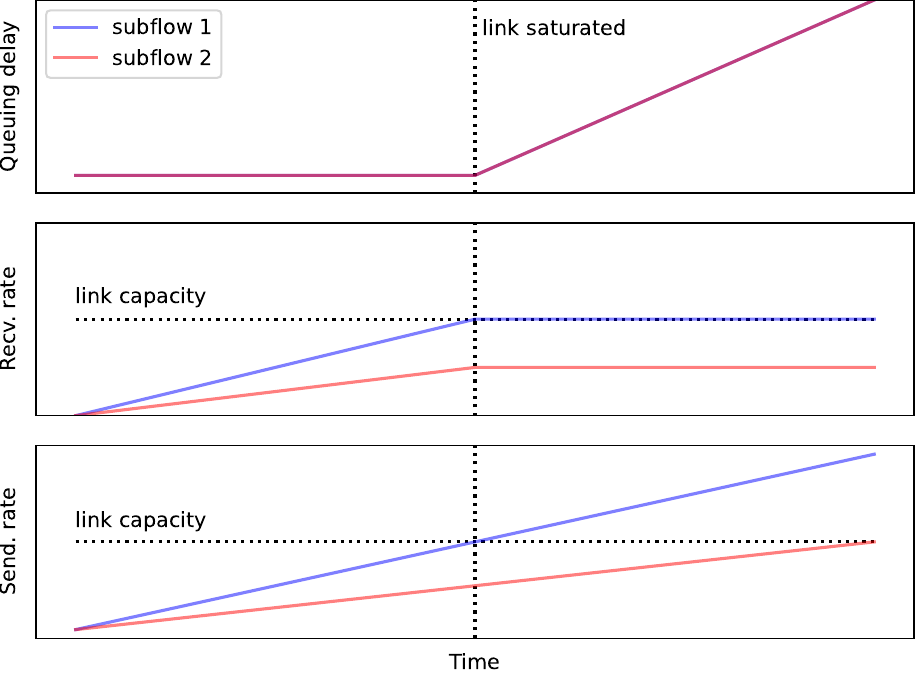}
\label{fig:no_fq_detail}}
\caption{In case there is \textbf{no fair queuing}, when the bottleneck link is congested, a queue builds up. Since both subflows share the same queue, they both see increased queuing delay.}
\label{fig:illustration_no_fq}
\end{figure}    

\section{Background}

When different applications send packets over the internet, one application can send more than the other and thus unfairly take a larger share of bandwidth. Also, a queue can form if one flow sends more than it should. 
This can result in unfairness and bad user experience. Several different approaches have been proposed to address this \cite{brown_future_2020,ware_beyond_2019}: One is to make sure every network flow is ``well-behaved'' (also known as TCP friendly). 
Another one is to enforce fairness at switches and routers, called ``fair queuing'' or ``flow queuing'' \cite{nagle_packet_1985}. 

While fair queuing was proposed decades ago, it only gained popularity in the last couple of years because of implementations in the Linux kernel \cite{dumazet_pkt_sched_2013,hoeiland-joergensen_flow_2018} and also in Apple's macOS. 
Applications can benefit from increased deployment of fair queuing: It makes sure that not the most aggressive one wins. 
It would be even better if applications could know if the connection they're sending on is managed by fair queuing. 
Then they could be sure that they can use their preferred congestion control mechanism, while not bothering or being bothered by other network flows. 
We proposed the first such approach in previous work \cite{bachl_detecting_2021} but our previous approach had some shortcomings upon which we improve in this paper. 

In our previous work we proposed a technique which determines the presence of fair queuing at flow startup, which works as follows: 
If fair queuing is successfully detected at flow startup, a congestion control is used which aims to keep queuing delay low (delay-based congestion control). 
While this delay-based congestion control achieves high throughput and low delay, similar to the Vegas congestion control algorithm \cite{brakmo_tcp_1995}, 
it is vulnerable to be outcompeted by other network flows sending more aggressively, such as \cite{cardwell_bbr_2016,dong_pcc_2015,ha_cubic_2008}.
This means that our delay-based congestion control performs well but only when it doesn't have to compete with other flows. Thus it is only used if fair queuing is detected. 
If the absence of fair queuing is detected, our previous approach uses a more aggressive congestion control (PCC \cite{dong_pcc_2015}), which can compete better with other, more aggressive network flows,
but doesn't keep delay as low as our delay-based congestion control. 

While our previous approach had high detection accuracy (98\%), it also had some limitations:
\begin{itemize}
    \item It would \textbf{only} detect fair queuing at \textbf{flow startup}. 
    But if the bottleneck link changes during a flow, it could be the case that the previous bottleneck had fair queuing while the new one doesn't. This wouldn't have been detected. 
    \item It would detect fair queuing only after filling the queue at the bottleneck completely, \textbf{causing packet loss}. 
\end{itemize}

We would like to have an approach which 
\begin{itemize}
    \item \textbf{continuously checks} for the presence or absence of fair queuing, not only at flow startup. 
    \item \textbf{doesn't cause packet loss} while trying to determine if there is fair queuing or not. 
    \item can be \textbf{transparently used} on top of any congestion control algorithm. 
    \item lets each application \textbf{choose how much delay} it allows in case fair queuing is detected. Allowing a higher queuing delay can result in higher throughput. 
\end{itemize} 

\section{Concept}

Our new approach presented here is a mechanism which performs measurements to determine if there is fair queuing. Our mechanism can be added on top of any existing congestion control algorithm. 
The approach of performing measurements in congestion control became popular in the last couple of years and was already followed by \cite{cardwell_bbr_2016,dong_pcc_2015,goyal_elasticity_2020,hayes_online_2020}.

The core concept of our approach, which we call \textbf{\textit{Tonopah}}, is that a network flow is separated into \textbf{two subflows}. 
One flow constantly sends more data (\textbf{dominant subflow}), while the other one sends less (\textbf{non-dominant subflow}). 
When the total sending rate reaches the link capacity and fair queuing is present, 
\textbf{fair queuing} is going to \textbf{limit} the throughput of the \textbf{dominant subflow} (\autoref{fig:illustration_fq}). 
Since its throughput is limited, a queue is going to build up and delay is going to increase. If the delay only increases for the dominant subflow but not for the non-dominant one, this means that there's fair queuing. 

On the other side, if there is no fair queuing, when the sending rate exceeds the capacity of the bottleneck link, a queue is going to build up and queuing delay is going to rise for \textbf{both} the dominant and the non-dominant subflow (\autoref{fig:illustration_no_fq}). 
In this case, Tonopah doesn't behave different from a network flow which doesn't have fair queuing detection. 
This can be a potential advantage which a Tonopah-enabled congestion control can have over other congestion control algorithms such as BBR, which has been shown to be unfair to other flows sometimes \cite{ware_modeling_2019,hock_experimental_2017}.

Thanks to the use of Multipath QUIC \cite{liu_multipath_2022} the application does not see any difference because all data are recombined to one flow by Multipath QUIC. 

An interesting aspect of fair queuing detection is that the presence of fair queuing can be interpreted as a \textit{congestion signal}: 
Only if there is congestion, fair queuing can be detected. If the link is underutilized, flows don't ``fight'' for bandwidth and thus fairness doesn't need to be enforced at the bottleneck link. 
Thus, in this paper, we define a new congestion signal: The presence of fair queuing. Other already known congestion signals are, for example: 
packet loss, queue length, change of queue length, Explicit Congestion Notification \cite{mathis_relentless_2009,hayes_revisiting_2011}. 

\subsection{Details}

The dominant subflow receives $\frac{2}{3}$ of the total allowed sending rate, while the non-dominant one receives the remaining $\frac{1}{3}$. However, this can be changed as long as it is made sure that one subflow gets a significantly larger share than the other one. 

After each round-trip time, Tonopah checks if the queuing delay of dominant subflow was on average larger than the one of the non-dominant subflow. As the threshold we use 5\,ms: If the average queuing delay in the last round-trip time was 5\,ms higher for the dominant subflow compared to the non-dominant one, fair queuing is detected. However, it is worth noting that each application can choose this threshold individually. For example, a cloud gaming application might choose a threshold of 5\,ms and thus prioritize low delay over high throughput. However, a video chat application might want to emphasize high bandwidth more and thus choose a threshold of 20\,ms to get maximum throughput. 

Both subflows are controlled by the same congestion window. The differences in sending rates are achieved by using pacing. 
The basis of the implementation of Tonopah is \textit{Picoquic}\footnote{\url{https://github.com/private-octopus/picoquic/tree/master/picoquic}}. 
Picoquic includes support for Multipath QUIC and pacing, which we need for the implementation of Tonopah. Also, Picoquic implements several congestion control algorithms. 
We base Tonopah on the NewReno implementation of Picquic. When fair queuing is detected, the congestion window is reduced by $\frac{1}{8}$. 
This worked well for our proof-of-concept implementation but one could also specify any other behavior in case fair queuing is detected or even do nothing at all. 

\section{Evaluation}

Tonopah was implemented on top of Picoquic in C. It was added on top of the NewReno congestion control algorithm but it could also be added on top of other 
congestion control algorithms as well. By basing our implementation on Picoquic, Tonopah also supports Explicit Congestion Notification. 

We evaluated Tonopah on an Apple MacBook Air from 2011 with an Intel Core i5-2557M CPU at 1.7\,GHz. 
By choosing old hardware for experimentation, we can show that our algorithm doesn't use excessive compute and can also run well on weak hardware. 

We emulate the network using \textit{mininet} and use Linux's \textit{pfifo} queuing discipline in the case there is no fair queuing and \textit{fq\_codel} and \textit{fq} for fair queuing. 
For the evaluation we choose a simple network in which the sender is connected to the receiver via a switch. 

All the source code and evaluation code of this paper is freely available (refer to the very top of this paper). 

In the following evaluations we always run each network flow for 90\,s. Tonopah measures whether there is fair queuing or not continuously. 
If there is fair queuing in the evaluation scenario and Tonopah detects fair queuing correctly for 80\,s and fails to detect it for 10\,s, we 
state the accuracy as $\frac{80}{90} = 89\%$. 
We evaluate the performance of Tonopah over a range of delays from 10 to 100\,ms and link capacity from 10 to 100\,Mbit/s in a grid-like fashion. 
For each combination of delay and link capacity, 10 experiment runs are performed and the empirical cumulative distribution functions and the average detection accuracy are shown. 

As the buffer size at the bottleneck we set the bandwidth delay product but at least 100 packets when no fair queuing (pfifo) is used. When using fq for fair queuing, we set the buffer size for each flow using the same rule. When using fair queuing with fq\_codel, we use it in its default configuration. 

\subsection{Detection accuracy}

\begin{figure}[h]
\centering
\includegraphics[width=\columnwidth]{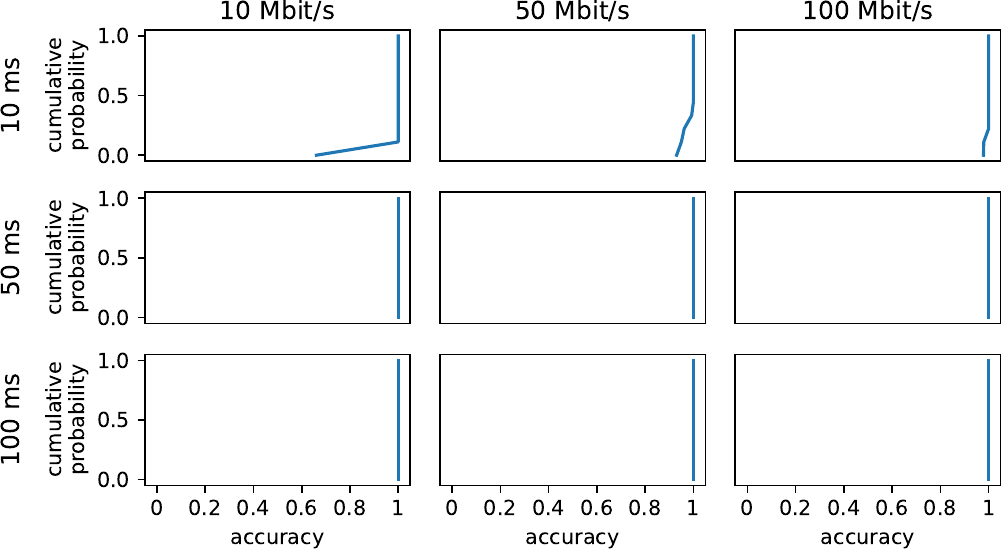}
\caption{Detection accuracy in case there's \textbf{no fair queuing} (pfifo). The overall accuracy is 99.4\%.}
\label{fig:no_fq}
\end{figure}

\begin{figure}[h]
\centering
\includegraphics[width=\columnwidth]{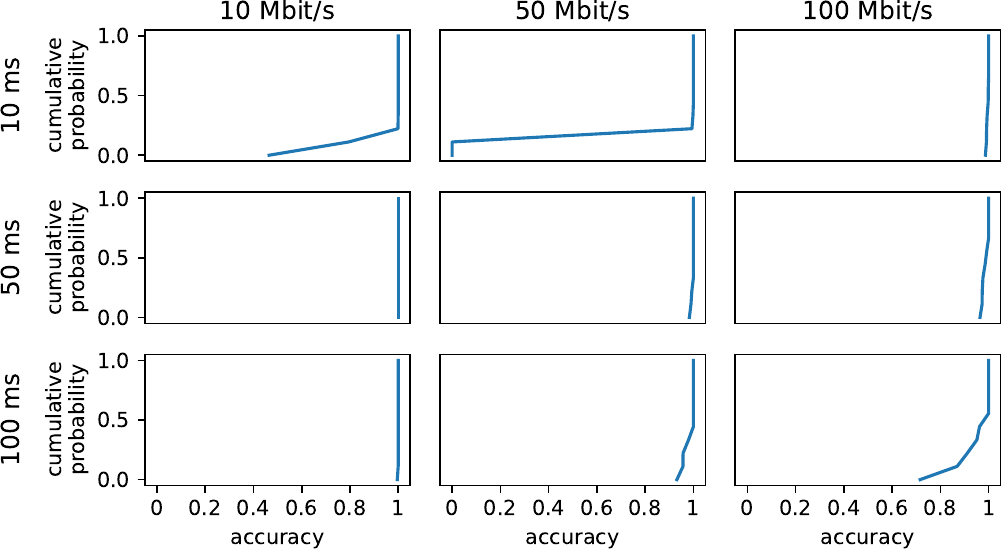}
\caption{Detection accuracy in case there is \textbf{fair queuing} (fq\_codel). The overall accuracy is 95.8\%.}
\label{fig:fq_codel}
\end{figure}    

\begin{figure}[h]
\centering
\includegraphics[width=\columnwidth]{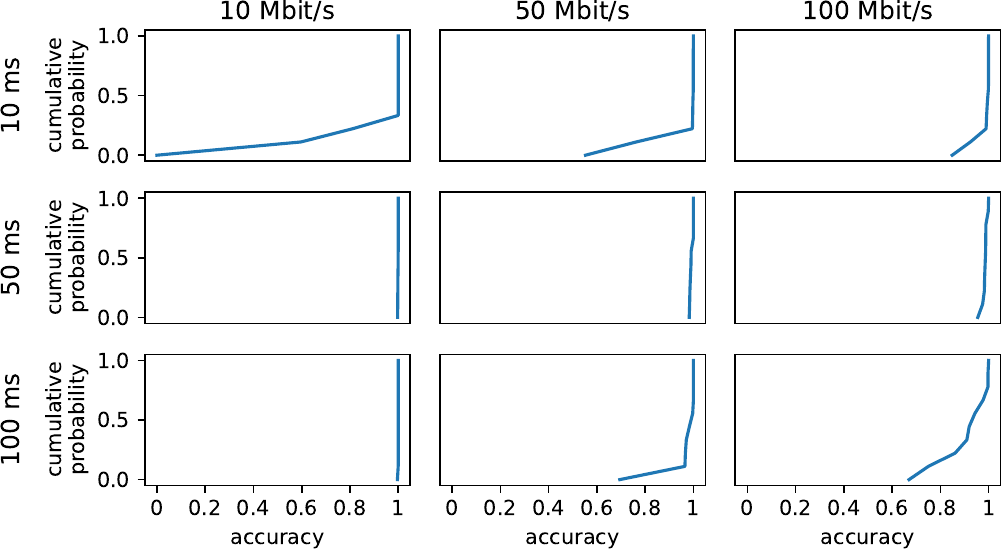}
\caption{Detection accuracy in case there is \textbf{fair queuing} (fq). The overall accuracy is 95.4\%.}
\label{fig:fq}
\end{figure}        

\begin{figure}[h]
\centering
\subfloat[\footnotesize pfifo]{\includegraphics[width=0.33\columnwidth]{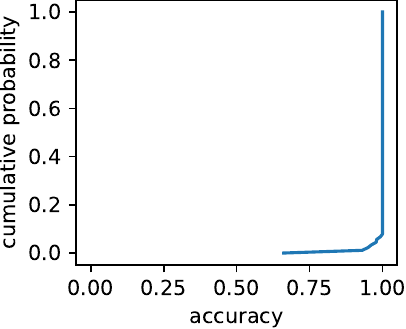}
\label{fig:ecdf_pfifo}}
\subfloat[\footnotesize fq\_codel]{\includegraphics[width=0.33\columnwidth]{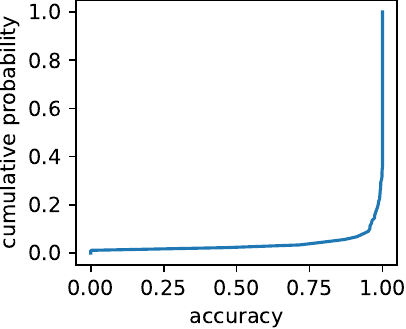}
\label{fig:ecdf_fq_codel}}
\subfloat[\footnotesize fq]{\includegraphics[width=0.33\columnwidth]{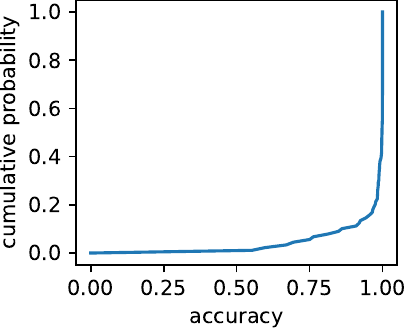}
\label{fig:ecdf_fq}}
\caption{Empirical cumulative distribution functions for detection of the abscence/presence of fair queuing.}
\label{fig:ecdf}
\end{figure}            

\autoref{fig:no_fq} shows that if there is no fair queuing, Tonopah detects this correctly virtually always. 
If there is fair queuing with fq\_codel (\autoref{fig:fq_codel}) or fq (\autoref{fig:fq}), Tonopah detects this correctly very often but not in every case. \autoref{fig:ecdf} shows the empirical cumulative distribution functions. 

\subsection{Cross-traffic}

\begin{figure}[h]
\centering
\includegraphics[width=\columnwidth]{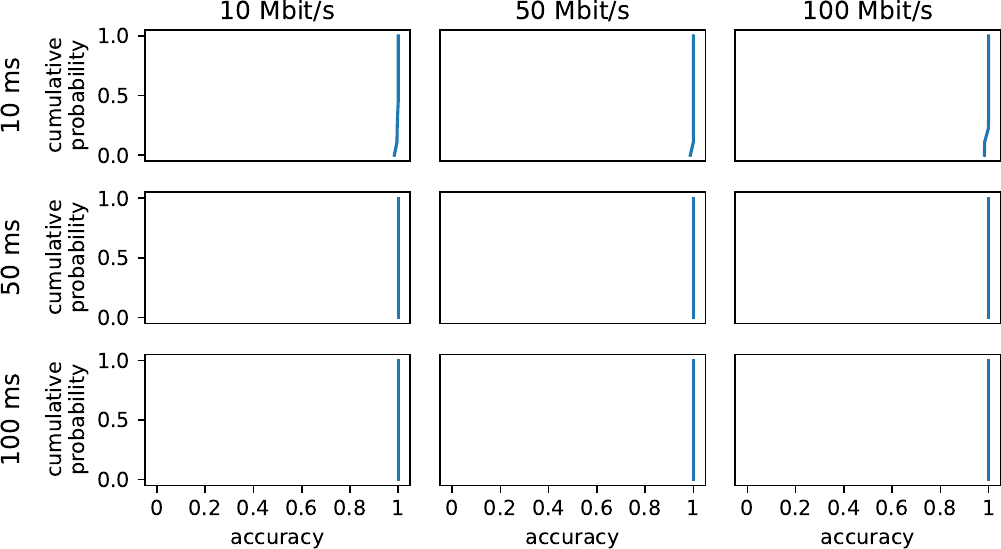}
\caption{Detection accuracy in case there's \textbf{no fair queuing} under the presence of \textbf{cross-traffic}. The overall accuracy is 99.9\%.}
\label{fig:no_fq_crosstraffic}
\end{figure}

\begin{figure}[h]
\centering
\includegraphics[width=\columnwidth]{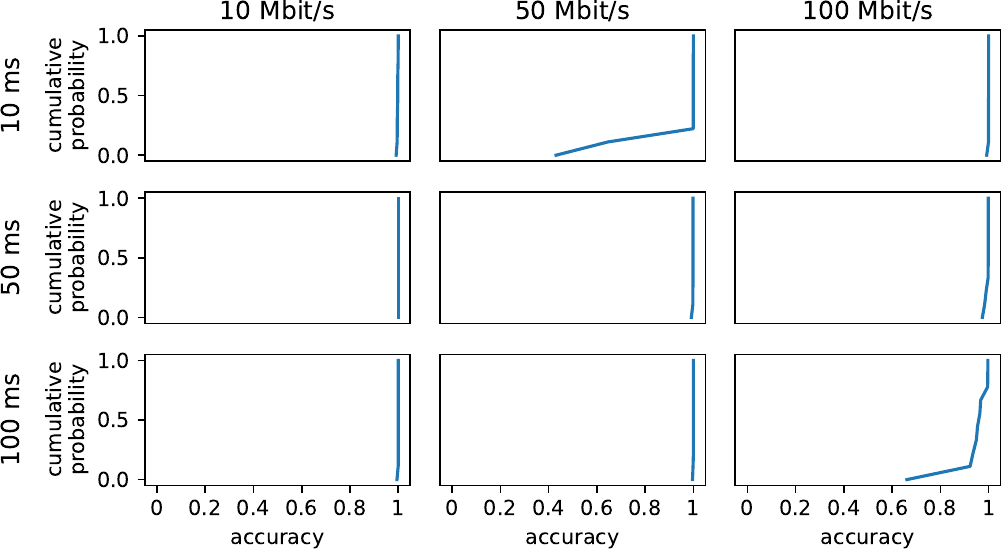}
\caption{Detection accuracy in case there is \textbf{fair queuing} (fq) under the presence of \textbf{cross-traffic}. The overall accuracy is 98.1\%.}
\label{fig:fq_crosstraffic}
\end{figure}    

\begin{figure}[h]
\centering
\subfloat[\footnotesize pfifo]{\includegraphics[width=0.33\columnwidth]{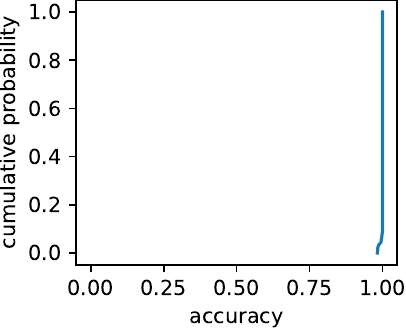}
\label{fig:ecdf_crosstraffic_pfifo}}
\subfloat[\footnotesize fq]{\includegraphics[width=0.33\columnwidth]{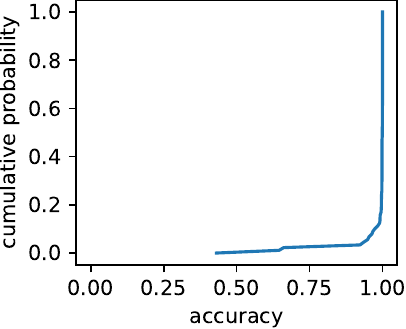}
\label{fig:ecdf_crosstraffic_fq}}
\caption{Empirical cumulative distribution functions for detection of the abscence/presence of fair queuing when there is \textbf{cross-traffic}.}
\label{fig:ecdf_crosstraffic}
\end{figure}                
    
To evaluate Tonopah's performance under cross-traffic, we use iperf3 with the NewReno congestion control. We start iperf 4\,s before Tonopah so that it has time to saturate the link. 
\autoref{fig:no_fq_crosstraffic} and \autoref{fig:fq_crosstraffic} show that also in the presence of cross-traffic, Tonpah's detection accuracy is virtually unchanged or even slightly better. We tried to use fq\_codel as in \autoref{fig:fq} but we encountered problems when using fq\_codel in combination with iperf3. Frequently, iperf3 would lose all packets, have a timeout after a couple of seconds and stop. Since we could not resolve this issue, we used fq instead of fq\_codel, for which this issue didn't occur. 

\autoref{fig:ecdf_crosstraffic} shows the empirical cumulative distribution functions when there is cross-traffic. 

\subsection{Throughput and Queuing Delay of Tonopah}

\begin{table}[h]
\begin{minipage}{\columnwidth}
\centering
\begin{tabular}{| r | r | r |}
\hline
& Link utilization & Queuing delay \\ \hline
NewReno & 95.6\% & 36.5 ms \\ \hline
Tonopah & 91.1\% & 3.3 ms \\ \hline
\end{tabular}
\caption{NewReno's link utilization is 105\% Tonopah's. 
But the \textbf{queuing delay} of NewReno is \textbf{1111\%} Tonopah's. 
This means that Tonopah can deliver the same or slightly less throughput while causing a lot less queuing delay. 
The differences in \textbf{queuing delay} are \textbf{highly significant} (p-value (Welch's t-test) $< 10^{-21}$). 
The experiment setup is the same as in \autoref{fig:fq}. }
\label{table:throughput_and_delay}
\end{minipage}
\end{table}    

While the goal of this paper is to implement a mechanism to detect fair queuing in an online fashion, 
we also show that fair queuing detection can be used to lower the queuing delay drastically while not impairing throughput. 
\autoref{table:throughput_and_delay} shows that Tonopah can lower queuing delay by more than $\frac{9}{10}$ while only achieving slightly lower throughput, when it is implemented on top on NewReno. 
However, we want to emphasize that Tonopah is not a congestion control algorithm but just a mechanism to measure fair queuing which can be used with any congestion control algorithm. 

\section{Discussion}

We showed that continuous fair queuing detection is feasible and can achieve a high detection accuracy for both fq\_codel and fq, although experiments showed that accuracy depends on the configuration of the queue manager, the link capacity, base delay etc. 
Further experiments could be done to finetune Tonopah and also experiment with other queue managers such as \textit{cake} \cite{hoiland-jorgensen_piece_2018}, \textit{PIE} \cite{pan_pie_2013}, \textit{cocoa} \cite{bachl_cocoa_2019} or \textit{LFQ} \cite{bachl_lfq_2020}. 

If one were to deploy Tonopah on the internet, one could implement an API, through which applications can choose how much delay they allow, in case fair queuing is detected. 

Since the subflow with the lower sending rate might experience lower queuing delay in case of fair queuing, it might be beneficial to include more important data in this subflow while the less urgent data can be sent in the subflow with the higher sending rate. This might, for example, be useful for cloud gaming: One could separate the video data into two parts: One part which includes a low resolution base video, which is transferred in the subflow with low queuing delay, while additional data, which increase the video quality, can be transferred in the subflow that might experience higher queuing delay. 

Another potentially fruitful line of research would be to integrate Tonopah into other congestion control algorithms such as BBR. 

\section*{Acknowledgements}

We thank Sebastian Moeller, who suggested to add CDF plots and to use fq\_codel with a target value of 5\,ms, which is the default on Linux. 

\bibliographystyle{ieeetr}
\bibliography{fair-queuing-aware-congestion-control}

\end{document}